\documentstyle[prd,aps,preprint,eqsecnum,epsfig]{revtex}
\newcommand{\be}{\begin{equation}}

\newcommand{\ee}{\end{equation}}
\newcommand{\bea}{\begin{eqnarray}}
\newcommand{\eea}{\end{eqnarray}}
\newcommand{\dst}{\displaystyle}
\newcommand{\bm}{\boldmath}
\newcommand{\fr}[2]{\frac{{\dst #1}}{{\dst #2}}}

\newcommand{\ggam}{\mbox{$\gamma\gamma\,$}}

\newcommand{\SM}{${\cal S} {\cal M}\;$}
\newcommand{\MSSM}{${\cal M}{\cal S}{\cal S} {\cal M}\;$}
\def\lsim{\mathrel{\rlap{\lower4pt\hbox{\hskip1pt$\sim$}}
    \raise1pt\hbox{$<$}}}         %less than or approx. symbol

\def\gsim{\mathrel{\rlap{\lower4pt\hbox{\hskip1pt$\sim$}}
    \raise1pt\hbox{$>$}}}         %greater than or approx. symbol

\title{Search for fourth generation quarks and leptons at the
Fermilab Tevatron and CERN Large Hadron Collider}

\author{I. F. Ginzburg,\thanks{E-mail: ginzburg@math.nsc.ru}}
\address{Sobolev Institute of Mathematics, Novosibirsk, 630090, Russia}
\author {I. P. Ivanov,\thanks{E-mail: i.ivanov@fz-juelich.de}}
\address{Sobolev Institute of Mathematics, Novosibirsk, 630090,
Russia and Forschungszentrum Juelich, Germany}
\author{A. Schiller\thanks{E-mail: schiller@tph204.physik.uni-leipzig.de}}
\address{Institut f\"ur Theoretische Physik and NTZ,
 Universit\"at Leipzig, D-04109 Leipzig, Germany}

\date{6 July 1999}
\tighten
\begin{document}

 \draft
  \maketitle
\vspace{7mm}

\begin{abstract}
If next generations of heavy quarks and leptons exist within the
\SM, they can manifest themselves in Higgs boson production at the
Tevatron and the LHC, before being actually observed. This
generation leads to an increase of the Higgs boson production
cross section via gluon fusion at hadron colliders by a factor
$6\div 9$.  So, the study of this process at the Tevatron and
LHC can finally fix the number of generations in the \SM. Using
the $WW^*$ Higgs boson decay channel, the studies at the
upgraded Tevatron will answer the question about the next
generation for mass values 135 GeV $\lsim M_H\lsim 190$ GeV.
Studying the $\tau\bar{\tau}$ channel we show its large
potential for the study of the Higgs boson at the LHC even in
the standard case of three generations.  At the Tevatron,
studies in this channel could explore the mass range 110 --- 140
GeV.
\end{abstract}

\pacs{PACS numbers: 12.60.-i,
14.60.Hi, 14.80.Bn}
\narrowtext
\section{Introduction}

The Standard Model (\SM) does not fix the number of fermion
generations.  So far it is not known why there is more than one
generation and what Law of Nature determines their number.

The studies on the $Z$ peak at LEP proved that there are exactly
three generations of quarks and leptons with light neutrinos.
However, the existence of next generations with heavy neutrinos is
not completely excluded. The $S$--parameter value obtained from modern
electroweak data \cite{pdg} makes the fourth generation disfavored
but only at the level of 2.5 $\sigma$ accuracy. Anyway, new quarks and
leptons should be heavier than the $t$--quark.

At a first glance, one cannot determine the number of
generations in the \SM before a direct observation of these new
particles.  Here we propose a simple way {\em how to determine
the number of such families before their direct discovery}
provided the simplest variant of the \SM (with one Higgs
doublet) is valid. The key is given by experiments probing the
Higgs boson $H$ production via gluon fusion at hadron
colliders (Tevatron and LHC). In accordance with modern
data, we have in mind that $M_H\gsim 95$ GeV \cite{90GeV}.  We
assume that the masses of new quarks $m_q\gg M_H$.

\begin{figure}[!htb]
 \centering
 \epsfig{file=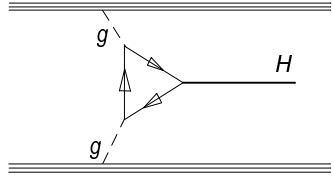,width=80mm}
 \caption{Dominant diagram for Higgs boson production in hadron collisions.}
 \label{fig1}
\end{figure}

The proposal stems out from well known facts (see e.g.
\cite{wilczek}). In $p\bar p$ or $pp$ collisions, the Higgs boson is
produced mainly via gluon fusion.  The production cross section is
proportional to the two--gluon decay width of the Higgs boson
$\Gamma(H\to gg)$. This width is induced by diagrams with quark loops
(see Fig.~\ref{fig1}).  The dominant contribution to $\Gamma(H\to
gg)$ comes from heavy quark loops. Indeed, the amplitude
corresponding to such a loop is $\propto{g_q\alpha_s}/{\mathrm{max}}
\{M_H,\,m_q\}$ (here $\alpha_s$ denotes the strong coupling
constant). The Yukawa coupling constant $g_q$ between the Higgs boson
and the quarks is $g_q=m_q/v$ (with $v=246$ GeV --- the Higgs boson
v.e.v.). At $m_q\gg M_H$, the quark contribution is finite and
$m_q$--independent. (The light quark contribution is suppressed as
$\sim(m_q/M_H)^2$.)

Therefore, in our discussion we restrict ourselves to the $t$--quark
and the possible fourth generation. Since the new fermion generation
contains two extra quarks, the amplitude of this decay increases by a
factor about 3 and the two--gluon width --- by a factor about 9 for a
light enough Higgs boson.

\section{The two--gluon width and gluon fusion}

In one--loop approximation the two--gluon decay width of the Higgs
boson can be written as
\be
\Gamma(H\to
gg)=\left(\fr{\alpha_s}{4\pi}\right)^2
\fr{M_H^3}{8\pi
v^2}|\Phi|^2\,,\quad \Phi=\sum\limits_q \Phi_q\,.
\ee
The quantity $\Phi$ is the sum of loop integrals $\Phi_q$
corresponding to different quarks $q$:
\bea
\Phi_q&=&-2r_q[1+(1-r_q)x^2(r_q)]\,,
\quad r_q=\fr{4m_q^2}{M_H^2}\,, \label{phi}\\
x(r)&=&\fr{\pi}{2}-\theta(r-1)\arctan\sqrt{r-1}\nonumber\\
&+& {\mathrm i} \,\theta(1-r)\ln\fr{1+\sqrt{1-r}}{\sqrt{r}}\,.\nonumber
\eea
With reasonable accuracy we consider only the $t$--quark and $N_q$
very heavy quarks from next generations (assuming that they are much
heavier than $H$). In this case
\be
 \Phi= -2r_t[1+(1-r_t)x^2(r_t)]
-\fr{4}{3}N_q\,.
 \label{phifin}
\ee
(If the heavy quark masses are comparable with $M_H$, all numbers have to
changed in a similar way as the effect of the $t$--quark mass seen in
Table~\ref{tab1} at $M_H>200$ GeV.)

In the following we denote by $\Gamma^k_{gg}$ and $\sigma^k$ the two
gluon width and the production cross section in a model with $k$
generations, for $k=4$ we have $N_q=2$. The discussed dependence of
the two--gluon width $\Gamma_{gg}^k$ on the Higgs boson mass $M_H$ in
one--loop approximation is shown in Table~\ref{tab1}.  (The effects
of radiative corrections are considered below.) Note that for a Higgs
boson mass close to the $t\bar{t}$ threshold, the relative
contribution of the top quark in Eq.~(\ref{phifin}) grows, which
leads to a decrease of the $\Gamma^4_{gg}/\Gamma^3_{gg}$ ratio.

The cross section for the Higgs boson
production in $pp$ ($p\bar{p}$) collisions is given as the
convolution of the cross section for the subprocess $gg\to H$
with the gluon structure functions $g(x,Q^2=W^2)$ ($W^2=x_1
x_2s$ is the total two--gluon c.m. energy squared). Since the
total Higgs boson width is less than 10 GeV for $M_H < 300$ GeV
and the distribution of colliding gluons is smooth, the standard
narrow--width approximation $\sigma_{gg\to H}=\pi^2\Gamma^k_{gg}
\delta(W^2-M_H^2)/(8M_H)$ is used with high precision.  With this form 
of $\sigma_{gg\to H}$, the cross
section of interest  reads
\be\dst\begin{array}{c}
\sigma(p\bar{p}\to H+\dots)\ =\\
\fr{\pi^2}{8s}\fr{\Gamma^k_{gg}}{M_H}
\int\limits^1_{M_H^2/s}\fr{dx_1}{x_1}
g(x_1,M_H^2)g(x_2=\fr{M_H^2}{x_1s},M_H^2)\,.
\end{array}\label{crsec}
\ee
The relative enhancement of the Higgs boson production due to the
fourth generation is simply
\be
\fr{\sigma^4({p\bar{p}\to H+\dots})}{\sigma^3({p\bar{p}\to
H+\dots})}=
\fr{\Gamma^4_{gg}}{\Gamma^3_{gg}}\,.\label{ratio}
\ee

The essential radiative corrections are of the following two sources:
\begin{itemize}
\item[i)] QCD corrections to the two--gluon width and the production
mechanism. Those corrections have been calculated for the \SM with
three generations in Ref.~\cite{qcd}. They enhance the Higgs boson
production cross section (\ref{crsec}) at the LHC by a factor
$K=1.5\div 1.7$. Varying the Higgs boson mass from 100 GeV to 1 TeV
the $K$--factor increases almost monotonously in that region. It changes
weakly for heavier quarks.
\item[ii)] Strong Yukawa interaction of the Higgs boson with heavy
quarks. The scale of these corrections is given by $g_q^2/(4\pi)^2=
m_q^2/(4\pi v)^2$. So, they can be neglected for heavy quark masses
$m_q<3$ TeV. (For example, at $m_q\leq 1$ TeV these corrections are
less than a few percent\ \cite{Meln}.) For still larger quarks masses
our results can be considered as a first rough estimate.
\end{itemize}

Therefore, the factor $K$ should be added in the cross section
(\ref{crsec}). In accordance with the results of Ref.~\cite{qcd}
and the previous discussion, we choose the value $K=1.5$ for both
the Tevatron and the LHC and at all Higgs boson masses. A more
precise value of $K$ would change our cross sections (and in
particular the ratio (\ref{ratio})) by less than 10 \%.

\begin{figure}[!htb]
 \vspace{-20mm}
 \centering
 \epsfig{file=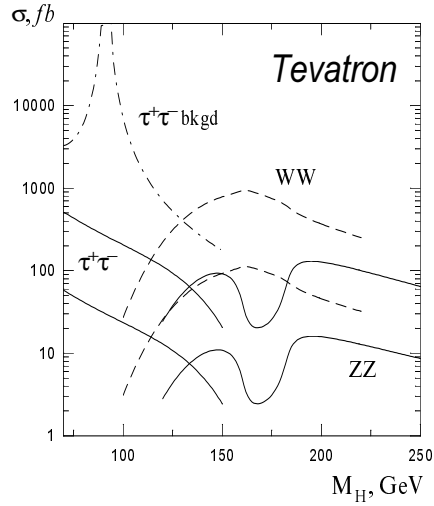,width=85mm,height=140mm}
 \vspace{-28mm}
 \caption{ The cross sections of the Higgs boson production for
  different decay channels calculated for the Tevatron.  The lower
  curves correspond to three generations, the upper ones --- to four
  generations. The $\tau\bar{\tau}$ background is shown by the
  dash--dotted line.}
 \label{fig2}
\end{figure}

\begin{figure}[!htb]
 \vspace{-40mm}
 \centering
 \epsfig{file=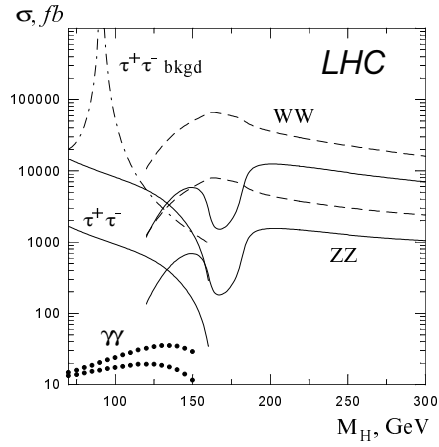,width=85mm}
 \vspace{-25mm}
 \caption{Same as Fig.~\protect\ref{fig2} for the LHC.}
 \label{fig3}
\end{figure}

Using the parameterizations of the gluon structure functions from
Ref.~\cite{strf}, we obtain the Higgs boson production cross sections
for different decay channels at the Tevatron (Fig.~\ref{fig2}) and
the LHC (Fig.\ \ref{fig3}) in the \SM with three or four generations.
To test the sensitivity to the parameterization of the gluon
structure functions we have calculated the same cross sections using
an alternative parameterization from Ref.~\cite{MRRS}. We observe that the
obtained cross sections differ weakly from those presented in the
Figures.

In the Higgs boson mass range 100 --- 140 GeV the  $\tau\bar{\tau}$
decay channel for the three family model has a branching ratio varying 
from 0.08 ($M_H=$ 100 GeV) to 0.04 ($M_H=$ 140 GeV).
For the four family model the corresponding branching ratio reduces from 
0.065   to 0.036.
At $M_H>150$ GeV the effect of the fourth
generation changes the observable branching ratios only
marginally. The main background for this channel is given by the
process $q\bar{q}\to Z(\gamma)\to\tau\bar{\tau}$. It is also
shown in Figs.~\ref{fig2} and \ref{fig3}.

The effect of the next generation can also be seen at the LHC
using the process $gg\to ZZ$ much above the Higgs resonance
position, at $m_q\gg M_{ZZ}\gg M_H$ \cite{Chan}. The main
contribution is given by the diagram with the $s$--channel Higgs
intermediate state.  The $ggH$ vertex entering in the measurable
cross section is $m_q$ independent (just as for the real Higgs
boson production) and the amplitude is $M_H$ independent at
$M_{ZZ}\gg M_H$. For Higgs boson masses $M_H<190$ GeV the
smaller cross section (far from the resonance) is compensated by
the opportunity to use the $ZZ$ final state with low background.
The resulting non--resonant effect is observable at the LHC
assuming a large enough luminosity integral.

\section{The two--photon width and the related cross section}

The fourth generation also modifies the
two--photon decay width of the Higgs boson. This decay
originates from similar loops with leptons, quarks and $W$
bosons. Here the $W$ boson and $t$--quark contributions (other
loops give negligible contributions) are of opposite sign and
the former dominates strongly at $M_H\ll 2m_t$. Thus the fourth
generation results\footnote{We included both heavy quarks and a
charged lepton, assuming that their masses are much larger than
$M_H$.} in a significant reduction of the two--photon width (see
Table~\ref{tab2}).

However, at $M_H\gsim 2m_t$ (where the $t$ quark contribution
exceeds the $W$ boson contribution) a new fermion generation
enhances the Higgs boson two--photon width again. This trend can
be also seen in Table~\ref{tab2}.  The QCD radiative corrections
are small in this case \cite{qcd}.

{}From the behaviour of the two--photon width we conclude that the
fourth generation would result in a strong (but destructive in a
wide range of masses $M_H$) effect for the Higgs boson
production at photon colliders (see Fig.~\ref{fig4}).

Concerning hadron colliders, this decay channel is of specific
interest if the Higgs boson will be discovered via the
$\gamma\gamma$ decay at the LHC. Note that at $M_H> 100$ GeV the
cross section of the process $gg\to H\to\ggam$, being
proportional to ${\mathrm{Br}}^k_{\gamma\gamma} \Gamma^k_{gg}$, increases
if the fourth generation exists (Table~\ref{tab2}). (Here
${\mathrm{Br}}_\phi$ is, as usual, the branching ratio for the decay
channel $\phi$.) The influence of an extra family becomes
stronger with increasing Higgs boson mass.

\section{Search strategies at the Tevatron and the LHC}

For a known luminosity integral ${\cal L}$ and detection
efficiency of a considered decay channel $\varepsilon$
we calculate below the significance of an effect   as the
ratio $S/\sqrt{B}$ with the
number of events for the  signal $S$ and the background $B$.
The number of events for the signal with $k$ generations is given by
$S_k=\varepsilon {\cal
{L}}\;\sigma^k_ {\mathrm{signal}}$, the corresponding number  
for the background by $B =
\varepsilon{\cal{ L}}\;\sigma_{\mathrm{bkgd}}$. Both $S$ and $B$ are
calculated with necessary cuts.
In our calculations we use the integrated
luminosity ${\cal {L}}=30$ fb$^{-1}$ for the upgraded Tevatron and
${\cal {L}}=100$ fb$^{-1}$ for the LHC.

\subsection{Different decay channels}

{\large\bm\bf The $b\bar{b}$ decay channel.} At $M_H<135$ GeV the
dominant decay channel is $b\bar b$. The signal to background ratio
$S/B$  
can be easily estimated since the main contribution to the
non--resonant background is given by the $b\bar{b}$ production in the
same gluon collisions.  Therefore, it is sufficient to compare the
non--resonant $b\bar b$ production with that occurring in the Higgs
boson decay ignoring the particular distribution of the gluon flux.
Using the just mentioned procedure to calculate the cross sections,
we find that the ratio $S/B<0.001$ even in the case of four
generations (more than $10^7$ events are necessary to obtain a significance
$S/\sqrt{B} > 3$). As a result, the $b\bar b$ channel cannot be used for
our problem.

{\large\bm\bf The $\tau\bar{\tau}$ decay channel.} In spite of the
relatively low branching ratio for this channel, the relative
value of the background is not so huge. The possibility to use the
$\tau\bar{\tau}$ decay channel for discovering the \MSSM
Higgs boson at the LHC and SSC was considered in
Ref.~\cite{Kunszt}. A more detailed study for the LHC has been performed 
recently in Ref.~\cite{Nik}.

When searching for a single $\tau$ lepton in $pp$ collisions,  
its leptonic decay channel should be used which has the branching
ratio $1/8.5$. Just this value of the detection efficiency was
achieved in the CDF experiment \cite{exp}. For the $\tau
\bar{\tau}$ pair the leptonic channel for one
$\tau$ and the hadronic jet channel for the another $\tau$
can be used \cite{Kunszt,Nik}. This procedure results in a branching ratio
$\tau\bar\tau \to l\nu + {\mathrm{jet}}$ of 0.09.  However, realizing
the presence of other background processes, various sources of
detection imperfection etc., we prefer to use a {\it twice lower
value} $\epsilon = 0.04$ to obtain cautious estimates. Using suitable cuts  
we hope that the resolution $\Delta M\approx
10$ GeV is achievable in the effective mass $M$ of the produced
$\tau\bar{\tau}$ system. In the calculations we use
 a cut--off $p_\bot>30$ GeV in the
transverse momenta of the produced $\tau$ and a rapidity cut--off  
$|\eta|<1.5$ for the Tevatron and $|\eta|<2.5$ for the
LHC.

We consider here the main source of a ``physical'' background --- the
process $q\bar{q}\to Z(\gamma)\to \tau\bar{\tau}$. The cut--off
in $p_\bot$ keeps 80 \% of the signal and 72 \% of the background
at $M_H=$ 100 GeV, 90 \% of  $S$ and 85 \% of $B$
at $M_H=$ 140 GeV.  The rapidity cuts leads to 88 --- 94 \% of the signal
and 76 --- 85 \% of the background (depending on the Higgs boson mass) and
improve the $S/\sqrt{B}$ ratio by   1 --- 2 \% for the Tevatron.  
At the LHC these cuts keep 84 --- 90 \% of the signal and
66 --- 71 \% of the background (depending on $M_H$) and
improve the $S/\sqrt{B}$ ratio by 4 --- 7 \%. From these estimates it follows
that both cuts do not cause any significant
improvement of the $S/\sqrt{B}$ ratio.  However, these cuts are still
useful because they cut off background processes of other
origin.  

Figs.~\ref{fig2},~\ref{fig3} and Table~\ref{tab3} show the
corresponding signal and background cross sections at the
Tevatron and the LHC (with a common $K$--factor $K=1.5$). The
last columns in that Table show the significance $S/\sqrt{B}$
defined above (both three and four generations are considered in
the case of the LHC) with a detection efficiency $\varepsilon=0.04$,
a luminosity integral 30 fb$^{-1}$ for the Tevatron and 100
fb$^{-1}$ for the LHC and the cuts mentioned above. One can see
that the effect of four generations will be observable at the
Tevatron in the Higgs boson mass range $M_H=$ 110 --- 140  GeV. 
At the LHC the effect will be huge for a
standard one--year luminosity 100 fb$^{-1}$.
Going from the Tevatron to the LHC the improvement of the ratio $S/\sqrt{B}$
is very natural, since with decreasing of $x\approx M_H/\sqrt{s}$ the
gluon structure function increases much faster than the quark
structure functions.

More detailed calculations can either improve these estimates 
or to make them worser.
To this end  additional backgrounds like
$t\bar{t}\to W^+W^-b\bar{b}\to\tau\bar{\tau} +\dots\to\dots$ should be
considered as it was made in Ref.~\cite{Nik}. Those backgrounds need detailed
simulations with some additional cuts (for example, the
anticoincidences with $b$--quarks can be useful). The relative
value of this background is essential at the LHC \cite{Nik}. 
At the Tevatron it corresponding value
 is much smaller  since the 
ratio $2m_t/\sqrt{s}$ is larger,  
resulting in a much
lower cross section for the $t\bar{t}$ pair production. Furthermore, in
our calculations we have averaged the cross sections over an
effective mass interval 10 GeV.  Detailed studies should
answer, whether this mass resolution is achievable or whether it it should be
improved.

{\large\bf\bm The $\mu^+\mu^-$ decay channel.} At the LHC
this channel can be used as an additional
opportunity to search for a new generation. The branching ratio of this channel
is $(m_\tau/m_\mu)^2\approx 283$ times smaller than
that for the $\tau{\bar{\tau}}$ channel. However, the mass resolution
here can be 1 GeV or even better \cite{Ritva}, the
signal is very clean, the detection
efficiency is close to 1 and the background is reduced by a factor $\gsim
10$. The last column in Table~\ref{tab3} for the LHC shows the 
$S_4(\mu)/\sqrt{B}$ value for the $\mu^+\mu^-$ channel with a mass 
resolution of 1 GeV. 

{\large\bf\bm The \ggam decay channel.} This channel is also
proposed for the Higgs boson study at the LHC \cite{gun}. The
accuracy needed to detect extra generations can be seen
from Table~\ref{tab2}.

{\large \bf \bm The $WW^*$ decay channel} is dominant at
$M_H>$ 135 GeV. The simulation of the process $gg\to H\to WW^*$
with background was performed within the \SM
with three generations in Ref.~\cite{HWW}. It was shown that the Higgs boson can
be excluded at 95 \% confidence level at 135~GeV$<M_H<190$~GeV
(which means that $S/\sqrt{B} \geq 2$) with a total luminosity
integral ${\cal{L}}=$ 30~fb$^{-1}$. Since the signal in our case increases
by a factor $p=8.2\div 8.5$   with an unchanged background, the
luminosity integral needed for the same significance is reduced
by $p^2$ to less than 0.5 fb$^{-1}$.      
Enhancing ${\cal{L}}$  by a factor $r$, the $S/\sqrt{B}$
ratio increases by $\sqrt{r}$. Therefore, in the case 
of four generations a luminosity integral of 3
fb$^{-1}$ is sufficient to see the Higgs boson in this mass
interval with $S/\sqrt{B}>5$.  An additional simulation is
necessary to find the lower bound of the accessible mass
interval for the luminosity integral 30 fb$^{-1}$. 

Fig.~\ref{fig3} shows that the opportunities at the LHC are
significantly richer.

{\large \bf \bm The $ZZ$ decay channel.} At $M_H >$ 190 GeV
this  channel is best suited for the Higgs boson search. The value of
the signal with reasonable kinematical limitations can be
estimated using the results of Ref. \cite{Chan} as input. For
the resonance production the signal is $\sim M_{ZZ}^2/(M_H
\Gamma_H)$ times larger than that calculated in Ref.~\cite{Chan}.
This factor is larger than 100 at $M_{ZZ}> M_H$ and $M_H <250$
GeV. In accordance with the numbers of Table~\ref{tab3}, the ratio
of products of gluon fluxes varies quickly with $M_H$: very roughly
it decreases by a factor of the order of 100 from the LHC to
the Tevatron in this mass region. Therefore, with a production
cross section being 8 times larger (Table~\ref{tab1}) the effect is
expected to be sizeable even at the Tevatron (run
III). The opportunity to observe the effect at lower values of
$M_H$ (the $ZZ^*$ channel) should be also explored.

\subsection{Different colliders}

{\large \bf Studies at the upgraded Tevatron.} Summarizing the
previous discussion, we conclude that the direct Higgs boson
production in the $\tau\bar{\tau}$ channel for the Higgs boson mass interval
110 --- 140 GeV and in the $WW^*$ channel for $M_H=$ 135 --- 190 
GeV will answer the question about the next generation for the
Higgs boson mass interval 110 --- 190 GeV. This statement is beyond doubts for
the $WW^*$ channel, and it should be tested in more
detail  for the $\tau\bar{\tau}$ channel. These studies
could reveal the existence of the fourth generation even if the
Higgs boson is not discovered in the associative production.

{\large \bf Studies at the LHC}.  At the LHC the Higgs boson is
expected to be visible in different channels depending on its
mass \cite{gun}.  In all cases, strong signals of the fourth
generation provide the opportunity to study the problem of new
heavy generations in the \SM even if the new particles are so
heavy that they cannot be directly produced. Using the
$\mu^+\mu^-$ channel with a possible mass resolution of about 1 GeV
and with a very clean signature is also possible despite the very
low branching ratio.

Besides, one can hope that the $\tau\bar{\tau}$ channel can
be used to observe the Higgs boson via gluon fusion even for the
\SM with three generations in the $M_H$ interval 100 --- 150 GeV.
Comparing the data in the $\tau\bar{\tau}$, $WW^*$, $ZZ^*$
($ZZ$) and $\gamma\gamma$ decay channels will be essential to
test the coupling constants of the Higgs boson with different
particles and thus verify the Higgs mechanism of the mass
generation in the \SM.  If $M_H<190$ GeV, the non--resonant $gg\to
H^*\to ZZ$ process \cite{Chan} will be a cross check of the
results obtained at the Tevatron.

\section{Concluding remarks}

Certainly, the presented numerical estimates are rough. Detailed
Monte Carlo simulations should take into account specific features of
the detectors (to account for the detection and triggering efficiency
in more details). They will show the exact regions of the Higgs boson
mass where the effect of the fourth generation could be seen in
particular channels at the Tevatron and the LHC.  For the $WW^*$
channel the simulation of Ref.~\cite{HWW} can be repeated using the
cross section with the fourth generation included.  For the
$\tau\bar\tau$ and $ZZ$ channels new simulations in a wide Higgs
boson mass interval are necessary.

\begin{figure}[!htb]
 \vspace{-20mm}
 \centering
 \epsfig{file=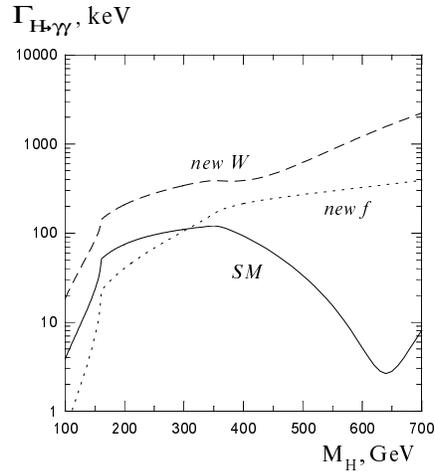,width=85mm}
 \vspace{-25mm}
 \caption{The cross section of the Higgs boson production at a
 photon collider (averaged over the photon spectrum) for different
 discovery modes: three generations (full line), four generations
 (dotted line), an extra charged vector boson (dashed line).}
 \label{fig4}
\end{figure}

The large effect in the gluon fusion discussed can be imitated by
other mechanisms, for example, in two Higgs doublet models or in
\MSSM at some values of the mixing angles $\beta$ and $\alpha$. In
this respect, additional measurements of the Higgs boson
production in $\gamma\gamma$ or $e\gamma$ collisions can help to solve our
problem. Indeed, adding to the \SM the fourth generation changes
strongly the relevant cross sections (see e.g. Fig. \ref{fig4}).
Fortunately, the parameters which imitate the effect of the
fourth generation in the gluon fusion do not give such an imitation
for the photon fusion ($H\gamma\gamma$ vertex) or the process
$e\gamma\to eH$ \cite{GIKr}. Therefore, the study of these
processes at photon colliders is very useful to obtain an
unambiguous conclusion about the existence of next heavy
generations.

\acknowledgments
We acknowledge very useful discussions with S.~Abdulin, R.~Kinnunen,
G.~Landsberg, M.~Krawczyk, N. Mokhov, A.~Nikitenko, E.~Richter--Was
and P.~Zerwas. I.F.G. is grateful to David Finley, Peter Lucas and
Hugh Montgomery for their hospitality to stay at FNAL. This work is
supported by grants of INTAS, RFBR 99-02-17211 and the Volkswagen
Stiftung I/72302.

\begin{table}[!htb]
 \begin{center}
  \begin{tabular}{cccc}
  $M_H$, GeV & $\Gamma^3_{gg}$, MeV & $\Gamma^4_{gg}$ MeV
  &$\Gamma^4_{g g}/\Gamma^3_{gg}\ $ \\
   \hline
 100 & 0.093 & 0.812 & 8.77 \\
 120 & 0.155 &  1.34 & 8.67 \\
 140 & 0.241 &  2.06 & 8.54 \\
 160 & 0.357 &  3.00 & 8.40 \\
 180 & 0.507 &  4.18 & 8.24 \\
 200 & 0.702 &  5.65 & 8.05 \\
 250 & 1.46  & 10.9  & 7.47 \\
 300 & 2.87  & 19.2  & 6.69 \\
 350 & 6.28  & 33.9  & 5.44 \\
  \end{tabular}\vspace{5mm}
  \caption{Two--gluon width for three and four generations as
         function of $M_H$.}
  \label{tab1}
 \end{center}
\end{table}

\begin{table}[!htb]
 \begin{center}
  \begin{tabular}{ccc}
  $M_H$, GeV &
  $\Gamma^4_{\gamma\gamma}/\Gamma^3_{\gamma\gamma}\quad \quad $ &
  $(\Gamma^4_{gg}Br^4_{\gamma\gamma})/
  (\Gamma^3_{gg}Br^3_{\gamma\gamma}) \quad \quad $ \\
  \hline
 100 & 0.158 & 1.12 \\
 120 & 0.193 & 1.31 \\
 140 & 0.250 & 1.78 \\
 160 & 0.403 & 3.32 \\
 180 & 0.485 & 4.00 \\
 200 & 0.539 & 4.34 \\
 250 & 0.710 & 5.30 \\
 300 & 0.960 & 6.50 \\
 350 & 1.36  & 7.40 \\
  \end{tabular}
  \vspace{5mm}
  \caption{Ratio of the two--photon widths
  $\Gamma^4_{\gamma\gamma}/\Gamma^3_{\gamma\gamma}$
  and of the cross section $gg \to H\to \gamma\gamma$ for four and
  three generations as function of $M_H$.}
  \label{tab2}
 \end{center}
\end{table}

\begin{table}[!htb]
 \begin{center}
  \begin{tabular}{ccccc}
  \multicolumn{5}{c}{\em Tevatron}\\ \hline \hline
  $M_H$, GeV &  3 generations & 4 generations &background
  &$S_4/\sqrt{B}$\\   \hline
  100 &  30 &   219   & 7400 & 2.75 \\
  110 &  24 &   166   & 1740 & 4.4  \\
  120 &  18 &   121   &  790 & 4.7  \\
  130 &  13 &    86.5 &  456 & 4.4  \\
  140 &   7 &    52   &  296 & 3.3  \\
  \end{tabular}
  \vspace{5mm}
  \begin{tabular}{ccccccc}
  \multicolumn{7}{c}{\em LHC}\\ \hline \hline
  $M_H$, GeV &  3 generations& 4 generations& background
  & $S_3/\sqrt{B}$ & $S_4/\sqrt{B}$&$S_4(\mu)/\sqrt{B}$\\ \hline
 100 & 1170   &  8270 & 35200 &  6.2 & 44   &  5.8 \\
 110 &  990   &  6330 &  8180 & 10.5 & 75   &  8.8 \\
 120 &  810   &  5550 &  3640 & 13.3 & 92   & 10   \\
 130 &  523   &  4210 &  2085 & 11.3 & 91   & 10.2 \\
 140 &  402   &  2840 &  1350 & 10.7 & 78   &  8.8 \\
 150 &  208   &  1610 &   946 &  6.7 & 51   &  5.9 \\
 160 &   31.5 &   257 &   690 &  1.0 &  8.6 &  1.1 \\
  \end{tabular}
  \vspace{5mm}
  \caption{Cross sections of the processes $pp\to \dots +gg\to H \to \tau
  \bar{\tau}$ for three and four generations and for the background   
  $pp\to q\bar{q}+\dots \to
  Z(\gamma)\to\tau\bar{\tau}$ at the Tevatron and the LHC
  (in {\em fb}). The $S_4(\mu)/\sqrt{B}$ for the $H\to\mu^+\mu^-$
  decay channel is  shown additionally.} 
  \label{tab3}
 \end{center}
\end{table}

\end{document}